\documentclass
[10pt,showpacs,prl,preprint,twocolumn,lengthcheck,nofootinbib]{revtex4}%
\usepackage{amsfonts}
\usepackage{amsmath}
\usepackage{amssymb}
\usepackage{graphicx}%
\providecommand{\U}[1]{\protect\rule{.1in}{.1in}}

\begin{document}
\preprint{ }
\title{2D Ising Model with non-local links - a study of non-locality}
\author{Yidun Wan}
\affiliation{Perimeter Institute for Theoretical Physics, Waterloo, ON N2L 2Y5, Canada}
\affiliation{Deptartment of Physics, University of Waterloo, Waterloo, ON N2L 3G1, Canada}

\begin{abstract}
Markopoulou and Smolin have argued that the low energy limit of LQG may suffer
from a conflict between locality, as defined by the connectivity of spin
networks, and an averaged notion of locality that emerges at low energy from a
superposition of spin network states. This raises the issue of how much
non-locality, relative to the coarse grained metric, can be tolerated in the
spin network graphs that contribute to the ground state. To address this
question we have been studying statistical mechanical systems on lattices
decorated randomly with non-local links. These turn out to be related to a
class of recently studied systems called small world networks. We show, in the
case of the 2D Ising model, that one major effect of non-local links is to
raise the Curie temperature. We report also on measurements of the spin-spin
correlation functions in this model and show, for the first time, the impact
of not only the amount of non-local links but also of their configuration on
correlation functions.

\end{abstract}
\volumeyear{2005}
\volumenumber{number}
\issuenumber{number}
\eid{identifier}
\maketitle
\date[dated: ]{Oct. 30, 2005}
\revised[Revised text]{: Jan. 26, 2006}

\received[Received text]{date}

\accepted[Accepted text]{date}

\published[Published text]{date}

In loop quantum gravity (LQG), states are described by spin networks, from
which non-local (NL) links can emerge due to the mismatch of micro-locality
and macro-locality\cite{Markopoulou2004}\cite{Smolin2005con}\cite{Smolin2005}.
In other words, two nodes which are local microscopically in the fundamental
graph can turn out to be macroscopically far away from each other and
connected by a NL link relative to the coarse grained metric. Ends of the NL
links carrying gauge field representations may be interpreted as gauge
particles, while the ends of those carrying no gauge fields can be some
unknown neutral particles and hence serve as a type of source of dark
matter\cite{Smolin2005con}\cite{Smolin2006}. It is found that the MOND force
law may\ also be recovered on a regular Euclidean lattice randomly decorated
with some NL links characterized by some distribution function $P(r)$, which
is the probability of having a NL link between two nodes separated by a
distance $r$ measured by the regular lattice metric\cite{Smolin2005}. Since NL
links correlate the fields at randomly chosen far away points without
violating the local conservation of averaged energy, they may explain the
non-locality in quantum physics and moreover, the origin of quantum physics
according to Nelson's formalism of stochastic quantum
theory\cite{Markopoulou2004}\cite{Smolin2005}.

To address these questions, we study the statistical behavior of\ 2D Ising
models, with randomly added NL links. Such a system is related to a class of
recently studied systems---small world networks (SWN), which are very useful
in describing many interesting social systems, such as railway network and
disease spreading\cite{Watts1998}\cite{Kleinberg2002}\cite{Newman1999}.
However, our approach and focus in this letter are different from those for
the SWN in that 1) we add a given number of NL links randomly to the regular
lattice with uniform distribution probability, and 2) we study and find the
effects of not only the amount of NL links but also the configuration of NL
links on the system's behavior, especially on correlation functions, which are
tightly related to the system's dynamics. To do so, we use Monte Carlo simulation.

\subsection{Simulation of 2D Ising Spin System with Random NL links}

We adopt Metropolis Monte Carlo method and apply periodic boundary condition.
We mainly simulate a $20\times20$ 2D square ferromagnetic Ising model randomly
decorated with NL links based on the uniform distribution probability. Note
that rewiring is not allowed and any two spins cannot be directly connected by
more than one link. We take the interaction coupling, $J=-1,$ and Boltzman
constant, $k_{B}=1,$ such that all the quantities presented are dimensionless.
The spin-spin correlation function $g\left(  r\right)  $ needs some
clarification. We study the pair correlation of the system as a function of
$r$, defined as the minimum number of links connecting one spin to another.
Because non-local links break the symmetry of the lattice, $g\left(  r\right)
$ should be obtained, for each $r$, by doing both ensemble average and the
average over the number of pairs of spins with the same distance $r$.
Mathematically, this is described by
\[
g\left(  r\right)  =\sum\limits_{\left(  i,j\right)  _{r}}\left[  \left\langle
\sigma_{i}\sigma_{j}\right\rangle _{r}-\left(  \left\langle \sigma
_{i}\right\rangle \left\langle \sigma_{j}\right\rangle \right)  _{r}\right]
/\#\text{pairs}\left(  i,j\right)  _{r},
\]
where the $\left\langle {}\right\rangle $ denotes ensemble average, the
subscript $r$ denotes the distance between two spins in the pair, and $\#$
means "the number of".

\subsection{Effects of the Amount of NL links}

Since we assume a uniform distribution probability for the NL links, the only
control we have on NL links is the number of them. Therefore, it is natural to
look at the effects of NL links on the system's behavior by varying their number.

\begin{figure}
[ptb]
\begin{center}
\includegraphics[
height=1.6786in,
width=2.0392in
]%
{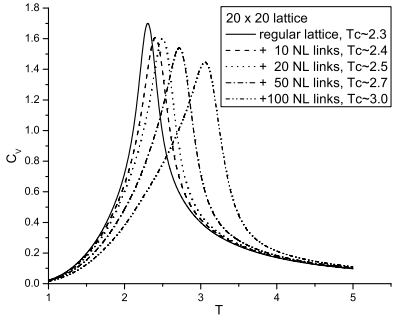}%
\caption{$C_{V}$-curves of respectively a regular $20\times20$ Ising spin
system, the same system with $10$, $20$, $50$, and $100$ non-local links. The
data of this plot is from Dr. Lixin Zhan with Wang-Landau method. }%
\label{fig_cv0_100nl}%
\end{center}
\end{figure}

Fig. (\ref{fig_cv0_100nl}) compares five curves, which correspond to the
specific heat of a $20\times20$ regular lattice with respectively $0$, $10$,
$20$, $50$ and $100$ NL links. One can see the increase of the critical
temperature with the number of NL links. This is so because, for a finite
system, adding NL links effectively increases the dimensionality of the
lattice and of course the average valence of the nodes. This argument can also
be verified by noticing the decrease of the peak height as the number of NL
links increases. Due to the finite scaling effect, the peak rises with the
system size, $L$, defined by $L^{D}=N$, where $N$ is the number of nodes of
the system and $D$ is the dimension. Now that $D$ increases with the amount of
NL links, but with $N$ fixed, $L$ must decrease and so does the peak height.
The last effect shown in the plot is that the half width of the $C_{V}$-curve
also increases with the amount of NL links. This is understood by noting that
1) the system with NL links has more states and 2) there are more non-frozen
states in the region of $T<T_{c}$ as $T_{c}$ increases with the amount of NL
links.%
Similar results are reported for SWNs on Ising model by Cai \textit{et.
al.}\cite{Cai2004}, by Herrero\cite{Herrero2002}, and by
Hastings\cite{Hastings2003}\footnote{Hastings proposed for a different model,
an approximated scaling relation between the increase of $T_{c}$ and the
probability $p$ of having NL links (long-range links in his work). His model
can be comparable to ours when $p\rightarrow0$. We have not found a close
match with his prediction, probably because we do not evolve sufficently large
lattices to obtain such a scaling behavior.}.%
\begin{figure}
[ptb]
\begin{center}
\includegraphics[
height=1.5342in,
width=2.0349in
]%
{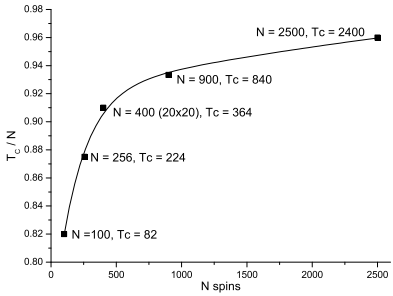}%
\caption{Plot of $T_{c}/N$ versus $N$, the system size, in the case of total
connectedness.}%
\label{fig_NvsTcBW}%
\end{center}
\end{figure}

One may ask if there is an upper limit of the critical temperature when we
keep adding NL links. Yes, there is a maximum $T_{c}$, which is the very
critical temperature in the extreme case when the system is totally connected.
In the stage of total connectedness, $T_{c}$ raises with the system size,
which is now equal to $N$. This is illustrated in Fig. (\ref{fig_NvsTcBW}),
where $T_{c}/N$ is asymptotic to unity in the limit of $N\rightarrow\infty$.
One can locate $T_{c}=364$ for a totally connected system of $400$ spins
(having $79800$ links), which corresponds to a $20\times20$ regular Ising
system with $79000$ NL links. This also implies that $T_{c}$, after a very
short linear range, increases slower and slower to approach the maximally
allowed $T_{c}$.

\begin{figure}
[ptb]
\begin{center}
\includegraphics[
height=1.593in,
width=2.0366in
]%
{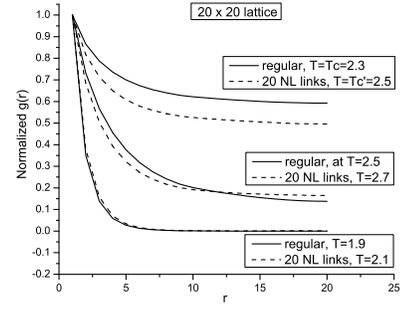}%
\caption{Comparison between the correlation function of the regular
$20\times20$ Ising model and that of the same lattice with $20$ non-local
links randomly distributed.}%
\label{fig_cf20nlcomparereg}%
\end{center}
\end{figure}

It is more interesting to see how the system's correlation function behaves in
the presence of NL links. Fig. (\ref{fig_cf20nlcomparereg}) shows three pairs
of correlation functions, each of which contains one for a regular lattice and
one for the same lattice with $20$ NL links. The upper pair compares the two
at their critical temperatures, one at $T_{c}=2.3$ and one at $T_{c}^{\prime
}=2.5$ due to NL links. One can see that both $g\left(  r\right)  $'s behave
similarly and indicate correlation lengths of the order of $L$, although with
different decay rates. By getting rid of the boundary effect, we have
$g\left(  r\right)  \propto r^{-0.2188}$ for the regular lattice, and
$g\left(  r\right)  \propto r^{-0.2996}$ for the one with $20$ NL links. In
the middle pair where each one is at a temperature of $0.2$ above its own
critical temperature, the two $g\left(  r\right)  $'s are closer to each other
and give rise to correlation lengths of the order of $L$, because the system
is finite and is in the vicinity of phase transition. However in this case,
one will see in the next section that there is some subtlety for the
correlation function of the system with NL links. In the lower pair where each
one is at a temperature of $0.4$ below its own critical temperature, the two
$g\left(  r\right)  $'s are almost the same and this is consistent with the
fact that the correlation length should be finite at such temperatures.
Therefore, even when the system is decorated with NL links, the number of
which is comparable to $L$, its correlation function is different but does not
deviate very much from that of the regular lattice.%

\subsection{Effects of the Configuration of NL links}

So far we just add, at the beginning of each run of the simulation, a certain
number of NL links on the lattice randomly based on the uniform distribution
probability, without any further control of their locations in the lattice.
Consequently, we may have different configurations of NL links on the lattice
in different runs of the simulation with all other parameters fixed. Now, let
us check if the configuration of NL links in the lattice has any impact on the
system's behavior.

\begin{figure}
[ptb]
\begin{center}
\includegraphics[
height=1.6319in,
width=2.0271in
]%
{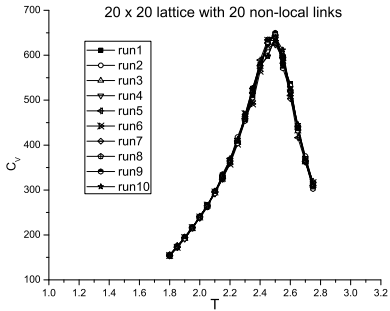}%
\caption{comparison of $C_{V}$ curves at $10$ runs of simulation with all
parameters the same except the configuration of non-local links. }%
\label{fig_cvcfg}%
\end{center}
\end{figure}

Actually, when the number of NL links is fixed, the $C_{V}$-curve is
self-averaging. This is verified by simulation. As an illustration, Fig.
(\ref{fig_cvcfg}) plots ten $C_{V}$-curves of ten runs (arbitrarily taken from
hundreds of runs) of the simulation with different NL link configuration but
with all other parameters the same. One can observe that all $C_{V}$-curves
coincide very well and peak at the same $T_{c}$. This suggests that for a
system, $C_{V}$ and hence $T_{c}$ are only sensitive to the total number of
links (both regular and NL links) on the lattice. The reason is that $C_{V}$
is a first order derivative of system's energy with respect to temperature and
the energy is only related the total number of links on the Ising lattice with
nearest neighbor interaction.%

\begin{figure}
[ptb]
\begin{center}
\includegraphics[
height=1.6241in,
width=2.1837in
]%
{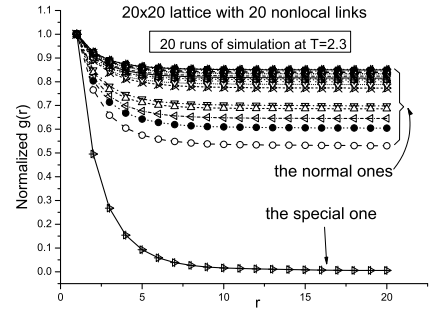}%
\caption{Comparison of correlation functions at $T=2.3$, which are obtained
from $20$ runs of simulation with same parameters, of a $20\times20$ Ising
lattice with $20$ non-local links.}%
\label{fig_l20nl20t2.3cf}%
\end{center}
\end{figure}

Nevertheless, if we look at the correlation function at temperatures not so
far from the critical temperature, configuration of NL links does effect.
Recall that we mentioned the subtlety of the correlation function of a
$20\times20$ lattice with $20$ NL links. This is illustrated in Fig.
(\ref{fig_l20nl20t2.3cf}), where we compare correlation functions at $T=2.3$
from $20$ runs (taken from hundreds of runs) of the simulation with different
NL link configuration but with the same parameters. The system is in the
vicinity of the phase transition, since $T_{c}\approx2.5$ in this case.
Nineteen $g\left(  r\right)  $'s of them are similar in that they all indicate
a correlation length of the order of $L$, with although different speed of
falling off. We call them normal ones in the sense that they behave similar to
the correlation function of the regular lattice at $T$ of the same distance
from the regular $T_{c}$. However, one of them, namely the special one, falls
off to zero rapidly and hence gives short correlation length of the system.
This is indeed due to a special configuration of the NL links on the lattice.
We show $20$ $g\left(  r\right)  $'s in this figure, because the ratio of the
special one to the normal one is roughly between $1/30$ and $1/20$.%

\begin{figure}
[ptb]
\begin{center}
\includegraphics[
height=1.6431in,
width=3.0286in
]%
{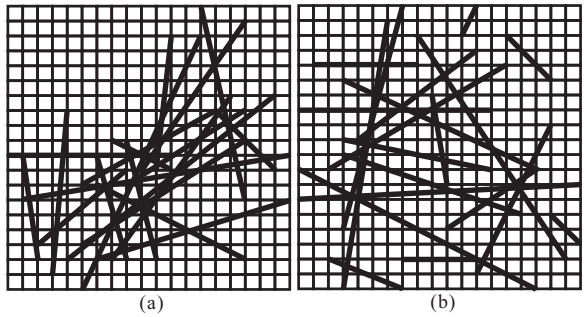}%
\caption{Two configurations of $20$ non-local links on a $20\times20$ lattice:
(a) configuration corresponding to the special correlation function in fig.
(\ref{fig_l20nl20t2.3cf}); (b) configuration corresponding to a normal
correlation function in fig. (\ref{fig_l20nl20t2.3cf}).}%
\label{fig_lattice_special_normal}%
\end{center}
\end{figure}

Fig. (\ref{fig_lattice_special_normal}) presents two different typical
configurations of the $20$ NL links on the $20\times20$ lattice. The
configuration corresponding to the special correlation function in Fig.
(\ref{fig_l20nl20t2.3cf}) is shown in Fig. (\ref{fig_lattice_special_normal}a)
(type A), where one can see that the whole upper left quarter of the lattice
has virtually no NL links. On the other hand, the configuration in Fig.
(\ref{fig_lattice_special_normal}b) (type B), which corresponds to a typical
one of the $19$ normal correlation functions, shows a much more evenly
distribution of NL links. By doing many simulations, we found that the length
$l$ of NL links, measured by the number of regular links taken from one end to
the other end of the NL link, are statistically different in the two cases.
For special ones, $l_{\min}\geq5$, and $l_{\mathrm{avg}}\geq10$, where
$l_{\min}$ is the minimum length and $l_{\mathrm{avg}}$ is the averaged $l$
over the number of configurations of the same type. But in the normal cases,
we have $2\leq l_{\min}\leq4$ and $l_{\mathrm{avg}}\leq10$. Based on these
observations, this phenomenon can be explained as follows. Since we choose the
uniform distribution probability of the NL links, in most of the cases one
should obtain configurations of NL links of type B; this explains why there
are more normal correlation functions than the special ones. In type B
configurations, NL links are short and very evenly distributed, so their
effect on the correlation function is well averaged out. As a consequence, the
system is correlated as there are no many NL links although $T_{c}$ is
different. Whereas, in type A configurations, NL links are longer and packed
in some region of the lattice. Provided this is so, the region where NL links
gather are effectively a sub-system of much higher dimension with of course a
much higher $T_{c}^{\prime}$, then the correlation function of this sub-system
behaves as at a temperature much below $T_{c}^{\prime}$ and gives rise to
short correlation length as it should. Moreover, the effect of NL links can
hardly be averaged out in such cases. Therefore, one sees the special
correlation functions.%

\subsection{Conclusions and Future Works}

We conclude that 1) the Curie temperature of the system increases with the
amount of NL links; 2) the Curie temperature is not sensitive to the
configuration of NL links; 3) normally, correlation functions of systems with
NL links behave similarly to those of regular lattices, with noticeable
difference in the decay rate; and 4) there exist some special cases where the
correlation functions of systems with NL links behave very differently from
regular ones due to some underlying special configurations of NL links on
lattices. Although it is hard to exclude the boundary effect due to the small
size of the systems we studied, the above results can still be valid for large
systems, especially when we are only interested in the physics in some portion
of the whole system. Nevertheless, to study large systems is of course
demanding and is one of our next steps. Conclusions one and two suggest some
manipulation of the interactions $J$ in the model, because $C_{V}$-curves and
$T_{c}$ will be subject to the configurations of NL links if we choose a
distance dependent interaction $J\left(  r\right)  $. Conclusions three and
four, on the other hand, suggest a modification on the distribution
probability $P\left(  r\right)  $ of the NL links. We see that correlation
functions suffer from the configuration of NL links, so we may get a special
type of correlation functions by controlling the configuration of NL links
through a well designed $P\left(  r\right)  $. For example, we may take a
$J\left(  r\right)  $, which has been studied by others and has physical
significance for a square spin lattice, and then look for a $P\left(
r\right)  $ with the above property. This would imply that a physical system
dominated by some interaction law can be equivalent to a system with non-local
links, which however obeys a much simpler interaction law. Some of these may
be done analytically. We can also extend the study to more complicated lattice
systems, or even to spin networks in the future.

\textit{Acknowledgements}---The author specially thanks Prof. Lee Smolin, the
author's supervisor, for his intuitions, discussions and comments. The author
thanks Dr. Lixin Zhan for checking some of the results. Thanks also go to
Mohammed Ansari, Doug Hoover, and Prof. Maya Paczuski and their useful
comments and discussions. The author would like to thank NSERC for the
financial support.%

\newif\ifabfull\abfulltrue%


\begin{thebibliography}{7}
\expandafter\ifx\csname natexlab\endcsname\relax\def\natexlab#1{#1}%
\fi\expandafter\ifx\csname bibnamefont\endcsname\relax\def\bibnamefont
#1{#1}\fi\expandafter\ifx\csname bibfnamefont\endcsname\relax\def
\bibfnamefont#1{#1}\fi\expandafter\ifx\csname citenamefont\endcsname\relax
\def\citenamefont#1{#1}\fi\expandafter\ifx\csname url\endcsname\relax\def
\url#1{\texttt{#1}}\fi\expandafter\ifx\csname urlprefix\endcsname\relax
\def\urlprefix{URL }\fi\providecommand{\bibinfo}[2]{#2}
\providecommand{\eprint}[2][]{\url{#2}}
\bibitem{Markopoulou2004}
\bibinfo{author}{\bibfnamefont{F.}~\bibnamefont{Markopoulou}} \bibnamefont
{and}
\bibinfo{author}{\bibfnamefont{L.}~\bibnamefont{Smolin}},
\bibinfo{journal}{Phys. Rev. D}\textbf{\bibinfo{volume}{70}},
\bibinfo{pages}{124029} (\bibinfo{year}{2004}),
\bibinfo{eprint}{see also: gr-qc/0311059}.
\bibitem{Smolin2005con}
\bibinfo{author}{\bibfnamefont{L.}~\bibnamefont{Smolin}},
\bibinfo{organization}{Loops'05},
\bibinfo{address}{Potsdam, Germany},
\bibinfo{month}{Oct.},
\bibinfo{year}{2005},
\bibinfo{eprint}{http://loops05.aei.mpg.de/index\_files/abstract\_smolin.html}%
.
\bibitem{Smolin2005}
\bibinfo{author}{\bibfnamefont{F.}~\bibnamefont{Markopoulou}} \bibnamefont
{and}
\bibinfo{author}{\bibfnamefont{L.}~\bibnamefont{Smolin}},
\bibinfo{journal}{In Preparation}.
\bibitem{Smolin2006}
\bibinfo{author}{\bibfnamefont{S. O.}~\bibnamefont{Bilson-Thompson}}
\bibnamefont{,} \bibinfo{author}{\bibfnamefont{F.}~\bibnamefont{Markopoulou}}
\bibnamefont{and} \bibinfo{author}{\bibfnamefont{L.}~\bibnamefont{Smolin}},
\bibinfo{journal}{In preparation}.
\bibitem{Watts1998}
\bibinfo{author}{\bibfnamefont{D.~J.} \bibnamefont{Watts}} \bibnamefont{and}
\bibinfo{author}{\bibfnamefont{S.~H.} \bibnamefont{Strogatz}},
\bibinfo{journal}{Nature} \textbf{\bibinfo{volume}{393}},
\bibinfo{pages}{440} (\bibinfo{year}{1998}).
\bibitem{Kleinberg2002}
\bibinfo{editor}{\bibfnamefont{T.~G.} \bibnamefont{Dietterich}},
\bibinfo{editor}{\bibfnamefont{S.}~\bibnamefont{Becker}}, \bibnamefont{and}
\bibinfo{editor}{\bibfnamefont{Z.}~\bibnamefont{Ghahramani}}, eds.,
\emph{\bibinfo{title}{Small-World Phenomena and the Dynamics of
Information}}, \bibinfo{organization}{NIPS} (\bibinfo{publisher}{MIT Press},
\bibinfo{address}{Cambridge, MA}, \bibinfo{year}{2002}).
\bibitem{Newman1999}
\bibinfo{author}{\bibfnamefont{M.~E.~J.} \bibnamefont{Newman}}
\bibnamefont{and} \bibinfo{author}{\bibfnamefont{D.~J.} \bibnamefont{Watts}},
\bibinfo{journal}{Phys. Rev. E} \textbf{\bibinfo{volume}{60}},
\bibinfo{pages}{7332} (\bibinfo{year}{1999}).
\bibitem{Cai2004}
\bibinfo{author}{\bibfnamefont{T.-Y.} \bibnamefont{Cai}} \bibnamefont{and}
\bibinfo{author}{\bibfnamefont{Z.-Y.} \bibnamefont{Li}},
\bibinfo{journal}{Intl. J. of Mod. Phys. B} \textbf{\bibinfo{volume}{18}},
\bibinfo{pages}{2575} (\bibinfo{year}{2004}).
\bibitem{Herrero2002}
\bibinfo{author}{\bibfnamefont{C.~P.} \bibnamefont{Herrero}},
\bibinfo{journal}{Phys. Rev. E} \textbf{\bibinfo{volume}{65}},
\bibinfo{pages}{066110} (\bibinfo{year}{2002}).
\bibitem{Hastings2003}
\bibinfo{author}{\bibfnamefont{M.~B.} \bibnamefont{Hastings}},
\bibinfo{journal}{Phys. Rev. Lett.} \textbf{\bibinfo{volume}{91}},
\bibinfo{pages}{098701-1}(\bibinfo{year}{2003}).
\end{thebibliography}
\end{document}